\documentclass[fleqn,10pt]{wlscirep}
\usepackage[utf8]{inputenc}
\usepackage[T1]{fontenc}
\usepackage{physics}
\usepackage{bm}
\usepackage{bbold}
\usepackage{color}
\usepackage{subcaption}
\usepackage{float}

\title{Topological Phase Transition in Superconductors with Mirror Symmetry}

\author[1]{Adam Lowe}
\author[2,+]{Miguel Ortuno}
\author[1,*,+]{Igor V Yurkevich}
\affil[1]{Nonlinearity and Complexity Research Group, School of Engineering \& Applied Science, Aston University, Birmingham B4 7ET, United Kingdom}
\affil[2]{Departamento de Física - CIOyN, Universidad de Murcia, Murcia 30.071, Spain}

\affil[*]{i.yurkevich@aston.ac.uk}

\affil[+]{these authors contributed equally to this work}

\begin{abstract}
	We provide analytical and numerical evidence that the attractive two-dimensional Kitaev model on a lattice with mirror symmetry demonstrates unusual 'intrinsic' phase at half filling. This phase emerges in the phase diagram at the boundary separating two topological superconductors with opposite Chern numbers and exists due to condensation of non-zero momentum Cooper pairs. Unlike Fulde-Ferrell-Larkin-Ovchinnikov superconductivity, the Cooper pairs momenta are lying along two lines in the Brillouin zone meaning simultaneous condensation of a continuum of Cooper pairs. 
\end{abstract}
\begin{document}
	
	\flushbottom
	\maketitle
	\thispagestyle{empty}
	
	\section*{Introduction}
	
Since the first development of the microscopic theory of superconductivity, the question of  nontrivial superconductivity in an isotropic Fermi-liquid corresponding to Cooper pairing with nonzero angular momentum has risen. The possibility of spontaneous breaking of rotational and time-reversal symmetries was demonstrated long ago \cite{Anderson}. In recent decades, a number of novel superconducting  materials with order parameters different from $s$-wave spin singlet  have  been  discovered. The  unconventional  superconductivity  has been observed in  high-Tc cuprate  superconductors,  ruthenates,  and  organic  materials.  It is believed that the pairing in these materials are caused by the electron-electron correlations different from the phonon-mediated point-like attraction.  Non-phononic nature of pairing is believed to be responsible for the observed spin  structure  and  orbital  symmetry  of  the  Cooper pairs. 

Superconductivity is a consequence of $U(1)$  gauge symmetry. Unconventional supersymmetry breaks also some other symmetries of the full symmetry group of the  normal state which includes usual space group operations (translations, rotations, and mirror reflections) and time  reversal  transformation. Unlike proximity induced symmetry breaking, 'intrinsic' mechanisms may also lead to to a spontaneously broken symmetries leading to nontrivial topology of quasiparticle excitations in the superconductors. The topology of an 'intrinsic' superconductor depends crucially on its order parameter which must be determined self-consistently to minimise a free energy. A change of a tuning parameter may result in a topological phase transition, for example in multicomponent superconductors. Configurations (functional dependence of the pairing field) minimising free energy may break spatial point group and time reversal symmetries. Such examples are known and the simplest nontrivial model is an attractive spinless fermions on a 2D square lattice. As long as the chemical potential is inside normal state band zone, below critical temperature $T_c$ there are two energetically equivalent superconducting states $p_x+i p_y$ and $p_x-i p_y$. The degeneracy can be lifted by any perturbation breaking mirror-reflection symmetry. These perturbations will also change the underlying superconducting order parameter to a conventional $p$-wave and topologically trivial superconducting state. The models with mirror-symmetry possess unavoidable degeneracy between $p_x\pm i p_y$ superconducting states. The Chern numbers of those states are $C=\pm 1$. Choosing a state, one may then change chemical potential and observe that the Chern number changes when the chemical potential goes through zero. The described above scenario is based on classification of topological superconductors with zero-momentum Cooper pairs.
	
We will show that the phase diagram of attractive spinless fermions on a square lattice contains an unusual superconducting state existing at the boundary where Chern number changes its sign (half filling at zero chemical potential). This state is different from usual superconductivity characterised by the emergence of off-diagonal long-range order, which results in a ground state where Cooper pairs have zero net momentum. It is similar to the so-called $\eta$-pairing, proposed earlier \cite{eta}, where electrons with momenta ${\bm k}$ and ${\bm\pi}-{\bm k}$ bind together leading to condensation of Cooper pairs with nonzero net momentum and spatially modulated superconductivity. This superconductivity is referred to as Fulde-Ferrell-Larkin-Ovchinnikov (FFLO) superconductivity \cite{FF},\cite{LO}. FFLO superconductivity was first proposed in a system with a significant Zeeman interaction, which shifts the Fermi surfaces for the up and down spins. It has been recently suggested that such state may emerge at one-third filling in the Kitaev-Heisenberg honeycomb model \cite{LeHur}. However, this exotic phase of matter has been observed only in a small number of systems so far.

\begin{figure}[h]
	\centering
	\begin{subfigure}{.5\textwidth}
		\centering
		\includegraphics[width=.6\linewidth]{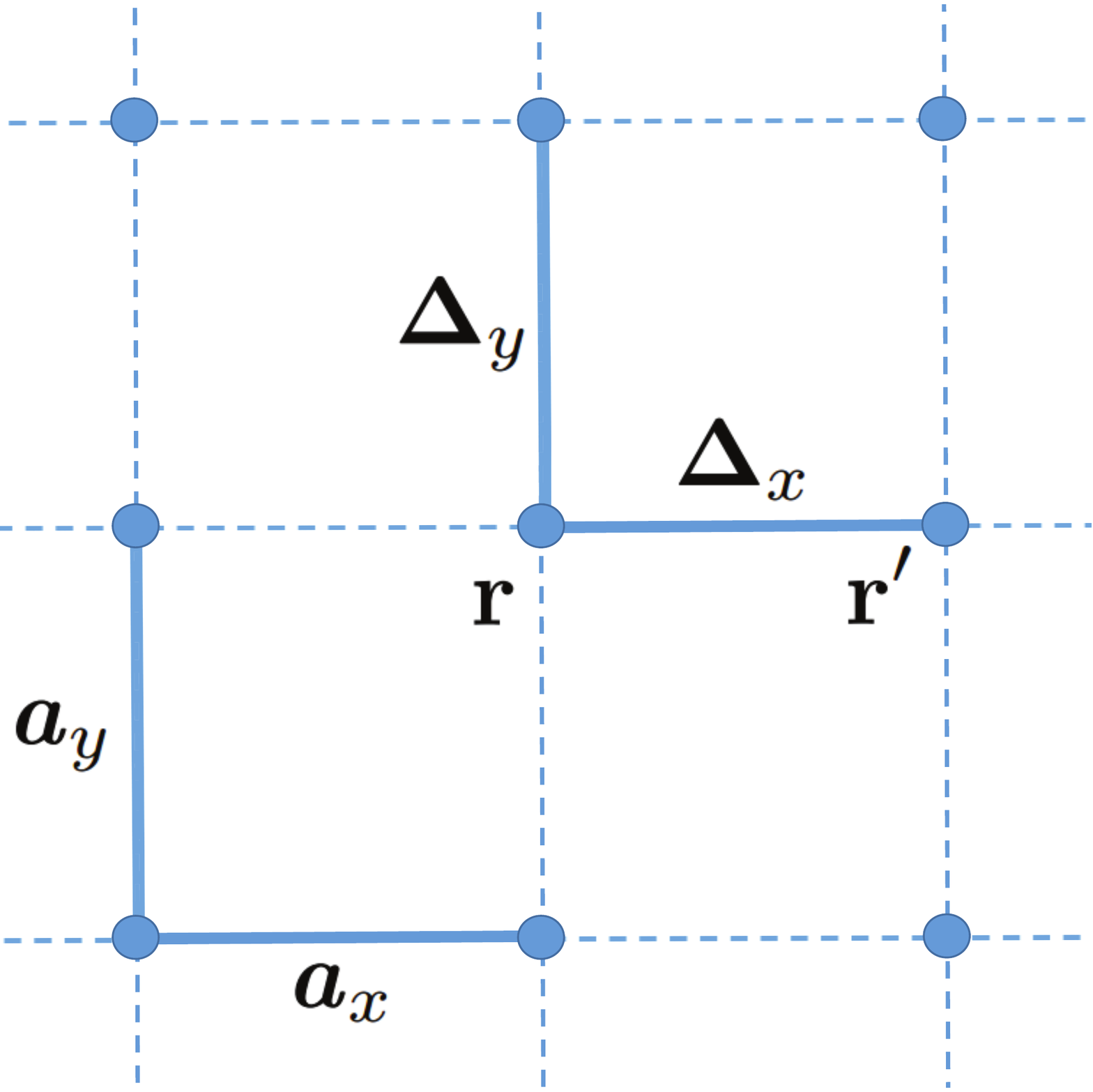}
		\captionof{figure}{Square lattice.}
		\label{fig:sl}
	\end{subfigure}%
	\begin{subfigure}{.5\textwidth}
		\centering
		\includegraphics[width=.6\linewidth]{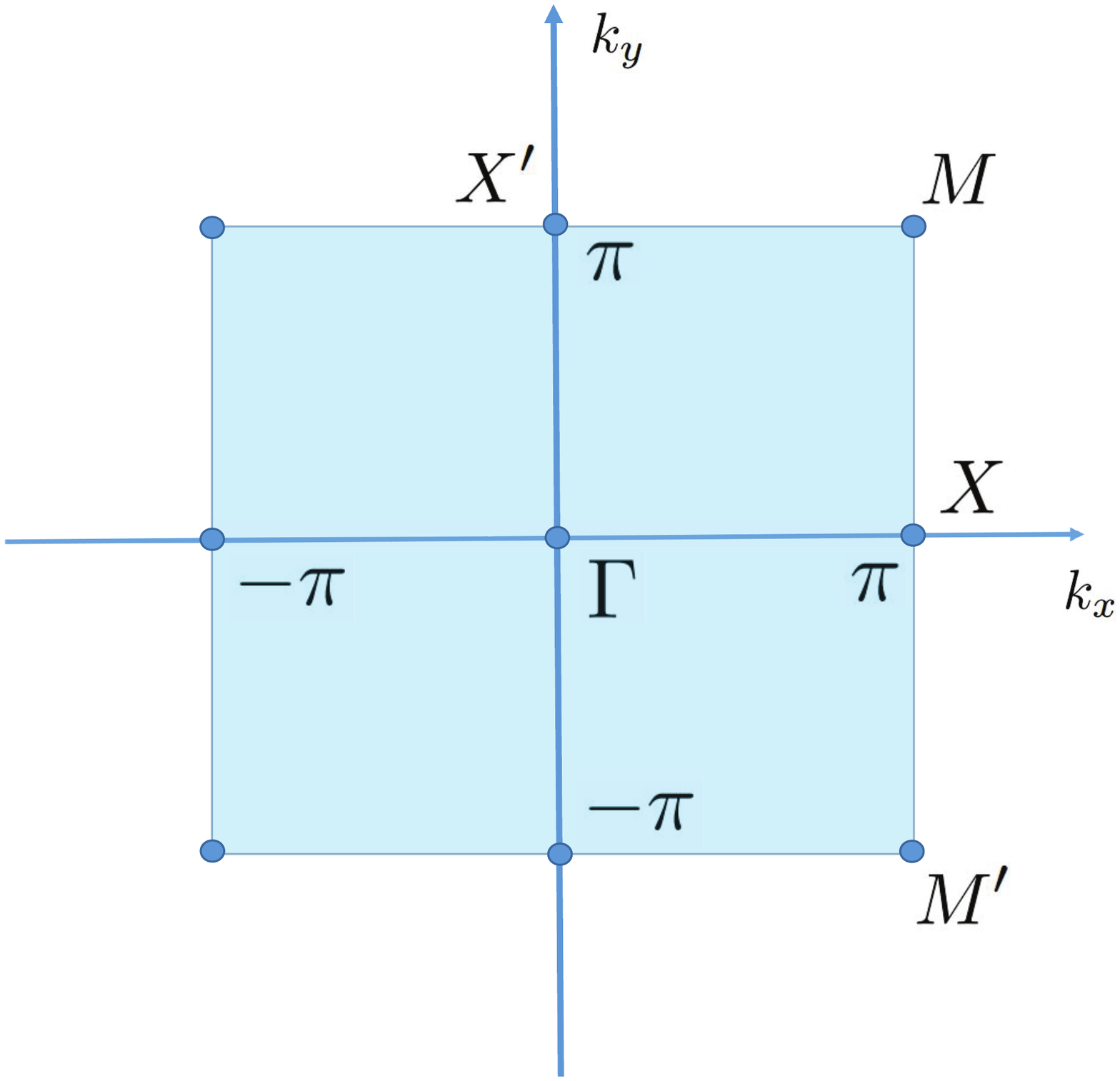}
		\captionof{figure}{Brillouin zone.}
		\label{fig:bz}
	\end{subfigure}%
\caption{(a) Square lattice with two lattice translation vectors ${\bm a}_x$ and ${\bm a}_y$ and pairing fields defined on vertical and horizontal links. (b) Brillouin zone and Fermi surface at half-filling. The gap function vanishes at all four inequivalent high-symmetry points $\Gamma$, $X$ and $M$.}
\end{figure}

In this paper, we restrict ourselves to the simplest model demonstrating nontrivial instability of the continuum of Cooper pairs with nonzero momenta.  We consider a 2D square lattice, Fig. {\ref{fig:bz}} with attraction between spinless fermions on adjacent lattice sites.The Hamiltonian of spinless electrons with attraction between nearest neighbours, $J> 0$, and the hopping, $\xi_{{\bf r}-{\bf r}'}$, has the form
\begin{equation}
{\hat H}=\sum_{<{\bm r},{\bm r}'>}\left[{\hat\psi}_{\bf r}^{\dagger}\,\xi_{{\bf r}-{\bf r}'}\,{\hat\psi}_{\bf r'}-\frac{J}{2}\,{\hat n}_{\bm r}{\hat n}_{{\bm r}'}\right]\,,\quad
{\hat n}_{\bm r}={\hat\psi}_{{\bf r}}^{\dagger}{\hat\psi}_{{\bf r}}
\end{equation}
where summation runs over adjacent lattice sites ${\bm r}$ and ${\bm r}'$.

\section*{Results}

The wave function of Cooper pairs emerging as the result of the Cooper instability is described by the pairing field naturally introduced by the Hubbard-Stratonovich transformation. This field acquires finite mean value below critical temperature and this value must be found from the minimisation of the free energy. Keeping only pairing field $\Delta_{\bm k}$ that couples electrons with opposite momentum (zero-momentum Cooper pairs), numerical results \cite{K} evidence that the gap function has the form  of $p$-wave order parameter, $\Delta_{\bm k}\sim\sin k_x\pm i\,\sin k_y$.
This order parameter describes unconventional superconductivity because it breaks time-reversal symmetry. Spontaneously broken time-reversal symmetry leads to non-trivial topological  properties of the superconductor. The BdG Hamiltonian with such order parameter is similar to QWZ model \cite{QWZ} characterised by the Chern number which, adapted to $p_x\pm i p_y$ superconducting phases, is known to be $C_{\pm}=\pm\,{\mathrm {sign}}\,\mu$.

For nonzero chemical potential ($|\mu|<2$) and away from half filling, $\mu=0$, one has two equivalent states describing topological superconducting phases, $p_x+i p_y$ and $p_x- i p_y$.  Below critical temperature, the sample goes into one of them (another spontaneous breaking). Once chosen (let us say it is $p_x+ip_y$ with the Chern number $C_+=\mathrm{sign}\mu$), this state may be manipulated by changing the chemical potential. The transition between two topologically nontrivial superconducting states occurs at zero chemical potential where the Chern number changes its sign. We would like to analyse the region where this transition takes place. We start with analysis of the superconducting instability and focus on differences between zero and nonzero chemical potential, i.e. between arbitrary and half filling occupation.

\begin{figure}[h]
	\centering	
	\begin{subfigure}{.49\textwidth}
		\includegraphics[width=\linewidth]{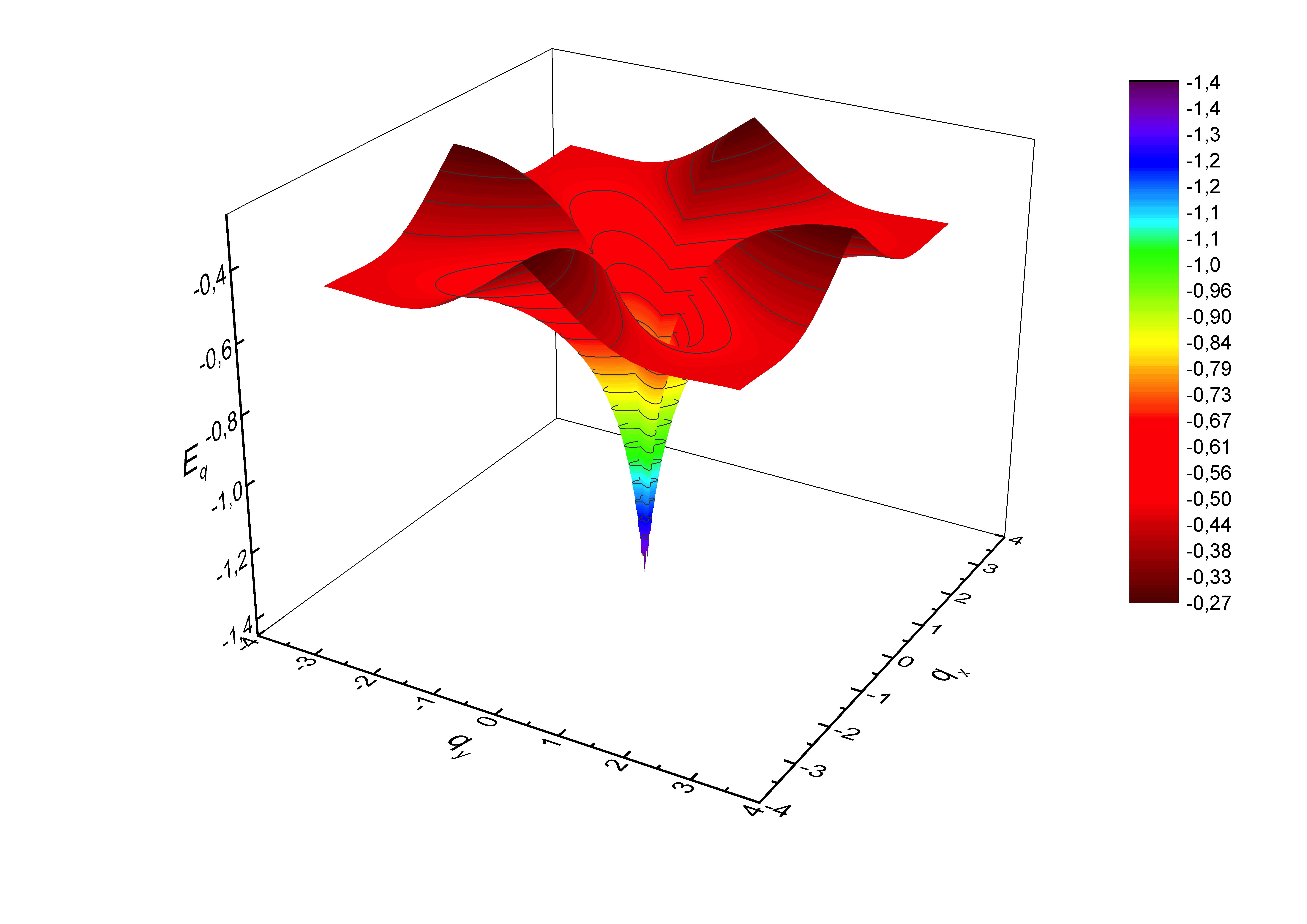}
		\caption{$\mu=-1$}
		\label{fig:mu1}
	\end{subfigure}
	\begin{subfigure}{.49\textwidth}
		\includegraphics[width=\linewidth]{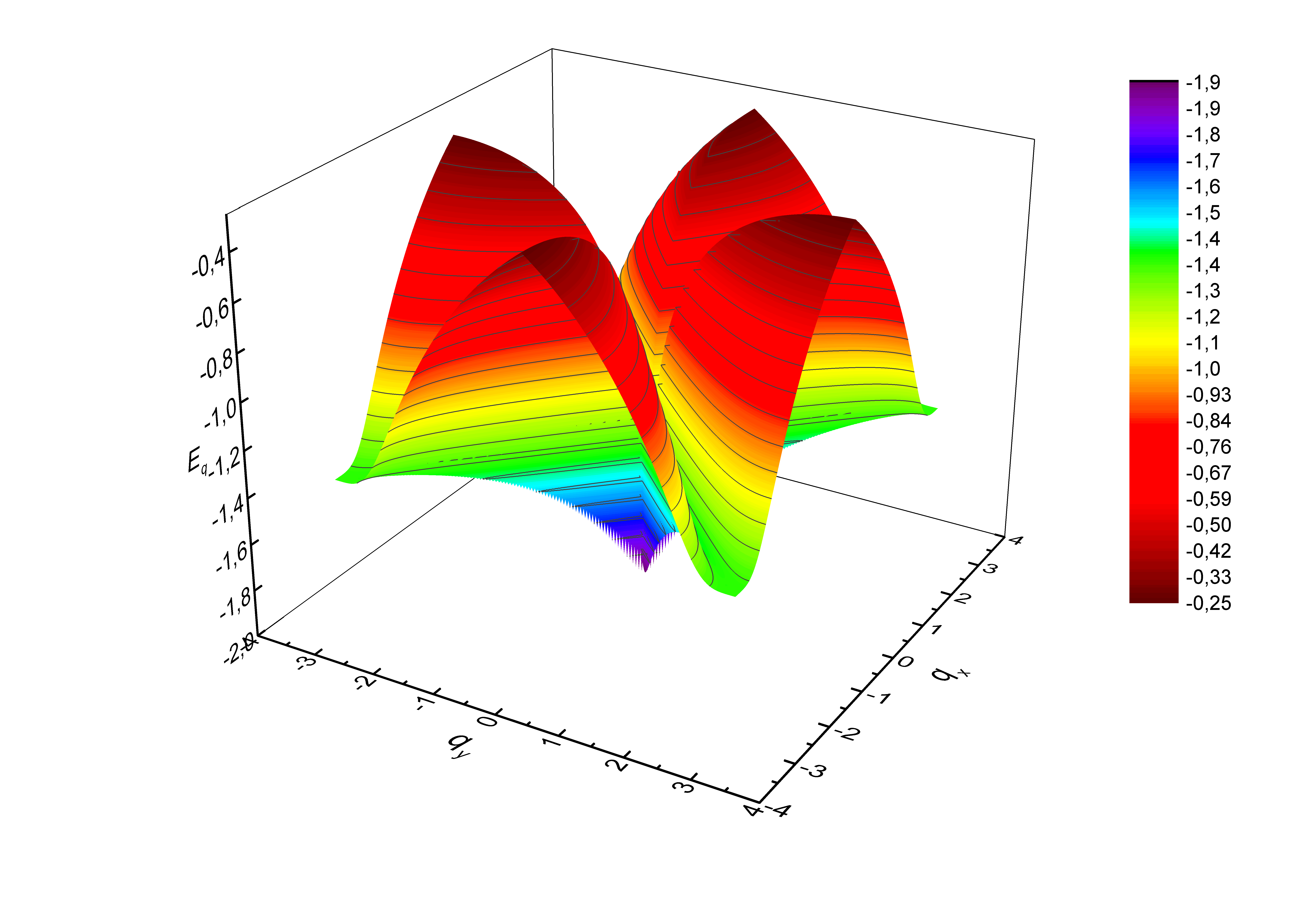}
		\caption{$\mu=0.1$}
		\label{fig:mu01}
	\end{subfigure}
\label{fig2}
	\caption{Energy of Cooper pairs with non-zero net momenta away from half-filling.}
\end{figure}

The superconducting instability in the vicinity of the transition is described by a quadratic in pairing field free energy
\begin{equation}\label{fe}
F_2=\sum_{\bm q}{\bm\Delta}^{\dagger}({\bm q})\left[J^{-1}\delta_{{\bm q},{\bm q}'}-{\hat\Pi}({\bm q})\right]{\bm\Delta}({\bm q})\,,\quad
{\bm\Delta}^{\mathrm T}=(\Delta_1, \Delta_2)\,,
\end{equation}
 where ${\hat\Pi}({\bm q})$ is the anomalous polarisation operator defined below in the $Methods$. The quadratic form may be diagonalised and the highest temperature $T$ at which the minimal eigenvalue of the operator $\left(J^{-1}{\mathbb 1}-{\hat\Pi}\right)$ becomes zero defines both critical temperature $T=T_c$ and the most relevant instability wavevector ${\bm q}={\bm q}_c$. The lowest eigenvalue $J^{-1}+E_{-}({\bm q})$ has the meaning of Cooper pair (bound state) energy. In the Figs.  (\ref{fig:mu1}) and (\ref{fig:mu01}), we present results of numerical simulations for the energies, $E_-({\bm q};\mu, T)$, of a Cooper pair with net momentum ${\bm q}$ for non-half filled band, $\mu\neq 0$. Numerics performed for different chemical potentials and temperatures supports conventional wisdom that minimum of the energy is reached at ${\bm q}=0$ which corresponds to homogeneous superconductivity, i.e. zero momentum Cooper pairs. It was proven earlier \cite{K}, one must expect development of standard (zero-momentum pairs) but unconventional $p_x\pm i p_y$ topological superconductivity with the Chern numbers $C_{\pm}=\pm\mathrm {sign}\mu$.
 
\begin{figure}[h]
		\centering
	\begin{subfigure}{0.49\textwidth}
		\includegraphics[width=\linewidth]{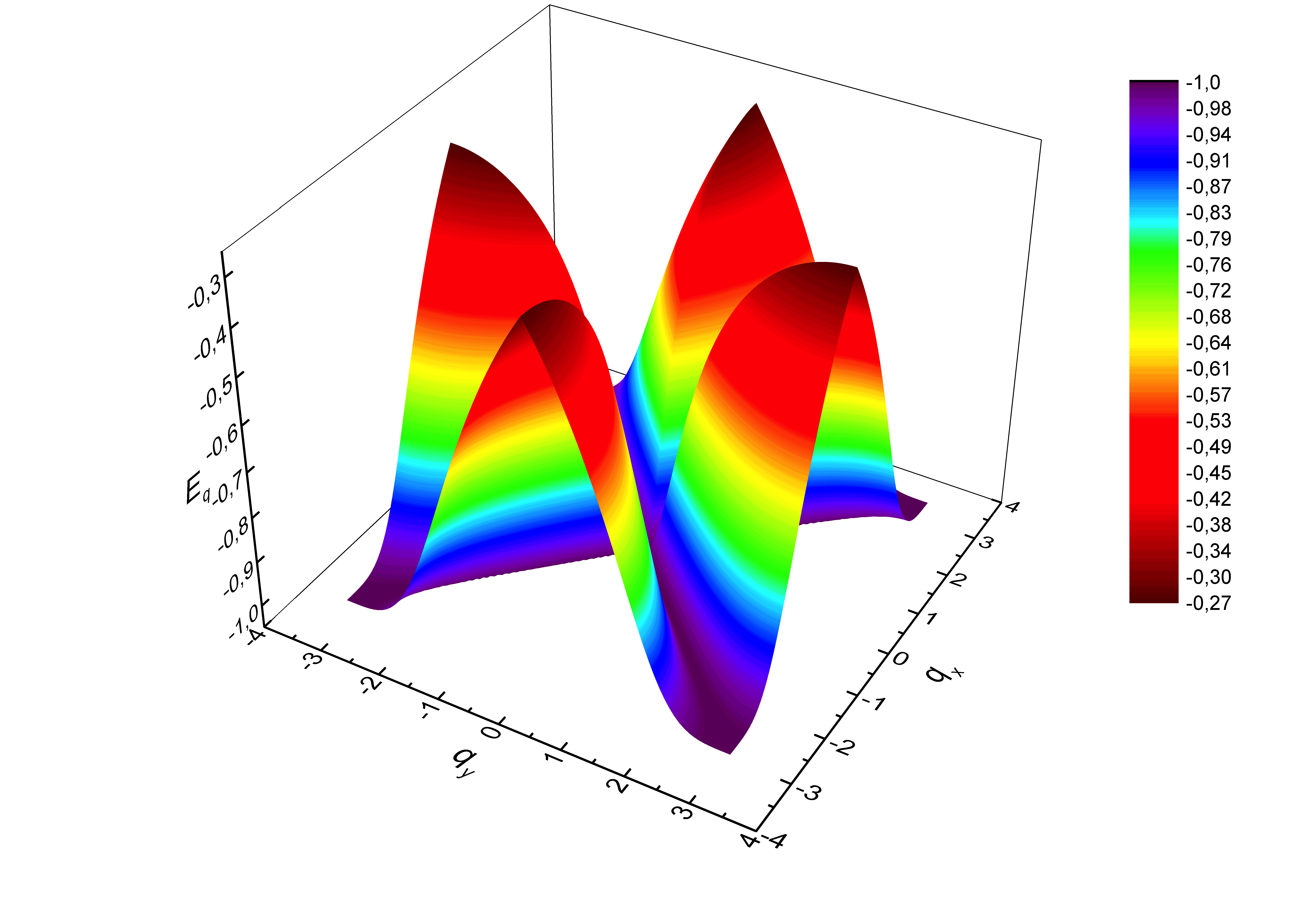}
		\caption{Eigenenergy of Cooper pairs with non-zero momenta at half filling $\mu=0$.}
		\label{fig:mu0}
	\end{subfigure}
	\begin{subfigure}{0.49\textwidth}
	\centering
	\includegraphics[width=\linewidth]{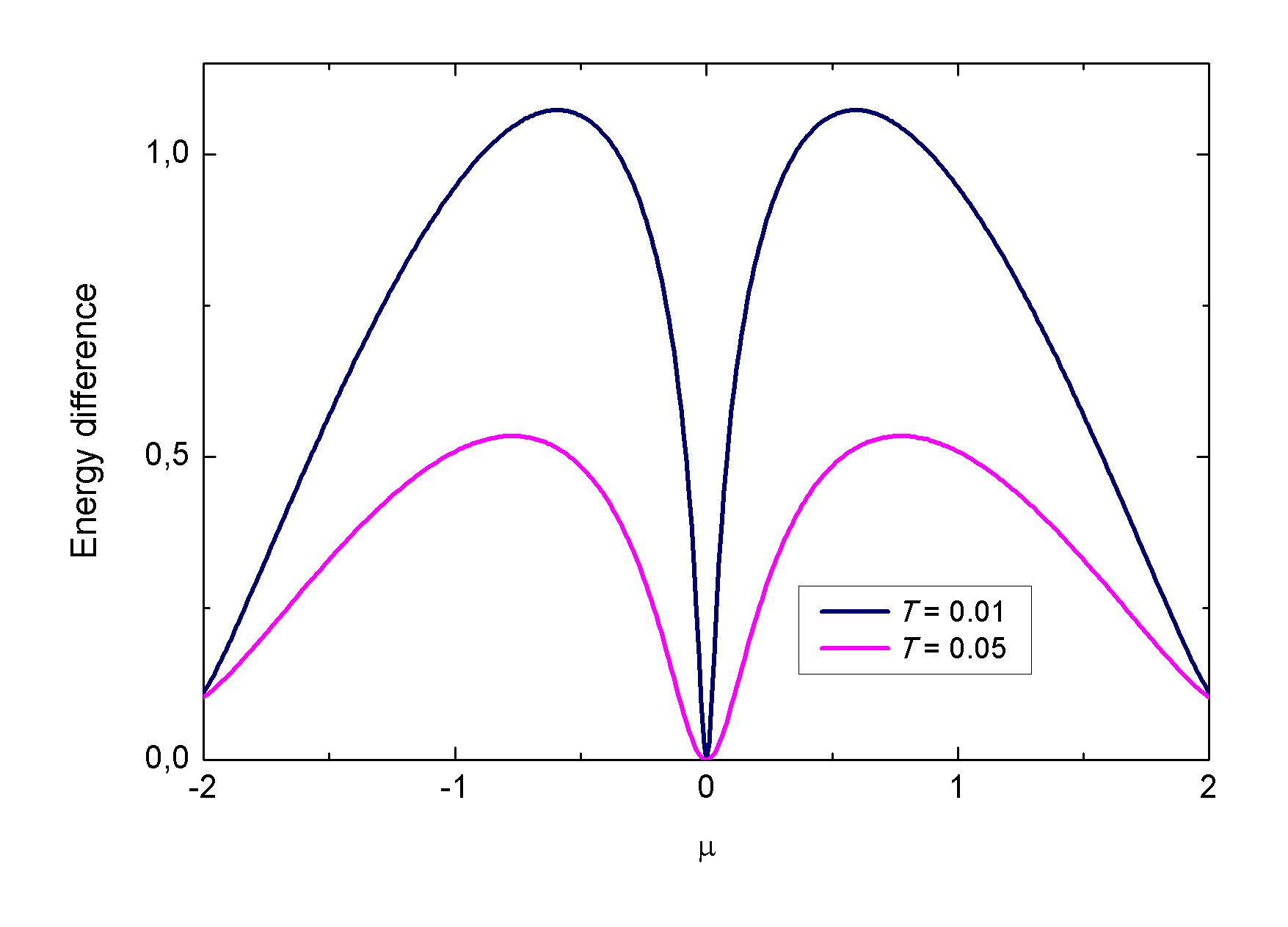}
	\caption{Parameter $\delta E(\mu)$ vanishing at the transition between two superconducting phases with different Chern numbers.}
	\label{fig:mu}
\end{subfigure}
\end{figure}

%\subsection*{Vicinity of superconducting transition}
The same analysis of quadratic instability described the free energy Eq. (\ref{fe}) at the half filling, $\mu=0$, shows that the minima of $E_-({\bm q};\mu, T)$ lie along the diagonals of the 'Brillouin zone' of the Cooper pairs (see Fig. (\ref{fig:mu0})). Indeed, it can be checked analytically that at $\mu=0$ the energy $E_-({\bm q})=const$ along the diagonals $q_x=\pm q_y$ and the derivative in the direction perpendicular to the diagonal vanishes.
	
This fact leads to the conclusion that minimum of the eigenvalue $J^{-1}+E_-({\bm q}; T_c,\mu=0)$ hits zero at some critical temperature along the diagonals signalling instabilities of a continuum of Cooper pairs with the momenta $q_x=\pm q_y$.  To visualize the degeneracy of the Cooper pairs energy, Figure (\ref{fig:mu}), we plot the parameter $\delta E(\mu)$ which is the difference between the energy of Cooper pairs at the corner of their Brillouin zone, ${\bm q}=(\pi,\pi)$, and at the origin, ${\bm q}=0$. This parameter is positive everywhere except at half-filling at $\mu=0$. Regions in the Figure (\ref{fig:mu}) with positive and negative chemical potentials, correspond to topological superconductors with condensed zero-momentum Cooper pairs and the Chern numbers of opposite sign. At the boundary, $\mu=0$, the parameter $\delta E(\mu)=0$ signalling simultaneous condensation of a continuum of Cooper pairs with momenta ${\bm q}=(1,\pm 1)\,q_x$ lying along diagonals  of the Cooper pairs Brillouin zone.

\section*{Discussion}
	
We have demonstrated that at half filling, exactly at the boundary between two phases where the 'intrinsic' superconductor becomes topologically nontrivial with non-zero Chern numbers, a different instability emerges due to continuous degeneracy of the Cooper pair energy. This degeneracy is responsible for simultaneous instability of a continuum of pairs with non-zero net momenta $q_x=\pm q_y$. The discovered instability is different from FFLO model where few momenta are singled out and modulate the order parameter.

To demonstrate our observation of a new superconducting transition not documented before, to the best of our knowledge, we have used the simplest possible model: 2D square lattice, attraction in a single triplet channel and no singlet channel interaction. It requires further investigation to decide how general our conclusion is and whether it is robust against perturbations like disorder.
	
\section*{Methods}
	
The interaction term can be written as a sum over the lattice vectors ${\bm a}$, Figure (\ref{fig:sl}),   and vectors ${\bm R}$ labelling inequivalent links:
\begin{equation}
S_{\mathrm int}=-\frac{J}{2}\sum_{<{\bm r}, {\bm r}'>}\,n_{\bm r}n_{\bm r'}
=-\frac{J}{2}\sum_{{\bm a}, {\bm R}}\,J_{\bm a}\,{\bar C}_{\bm a}({\bm R}) C_{\bm a}({\bm R})\,,\quad
C_{\bm a}({\bm R})={\psi}_{{\bm R}-{\bm a}/2}\psi_{{\bm R}+{\bm a}/2}\,,
\end{equation}
The Hubbard-Stratonovich transformation decouples the interaction,
\begin{equation}  
S  =S_0+\frac{1}{2}\sum_{{\bm a},{\bm R}}\left[\frac{|\Delta_{\bm a}({\bm R})|^2}{J}+\Delta_{\bm a}({\bm R}){\bar C}_{\bm a}({\bm R})+{\bar \Delta}_{\bm a}({\bm R}) C_{\bm a}({\bm R})\right],
\end{equation}
thus introducing the pairing field which is defined on the links. Fermion anticommutation requires this field to be an odd function of the lattice vectors, $\Delta_{\bm a}({\bm R})=-\Delta_{-\bm a}({\bm R})$. The coupling between fermions and pairing field can be written in the momentum representation:
\begin{equation}
\sum_{{\bm a}, {\bm R}} \Delta_{\bm a}({\bm R})\,{\bar C}_{\bm a}({\bm R})=\sum_{{\bf k},{\bm q}}\,\Delta_{\bm k}({\bm q})\,{\bar C}_{\bm k}({\bm q})\,,\quad
C_{\bm k}({\bm q})=\psi_{-{\bm k}+{\bm q}/2}\psi_{{\bm k}+{\bm q}/2}\,,\quad
\Delta_{\bm k}({\bm q})=\frac{1}{N}\sum_{{\bm a}, {\bm R}}\Delta_{\bm a}({\bm R})\,e^{-i{\bm q}{\bm R}-i{\bm k}{\bm a}}\,.
\end{equation}
Superconducting field $\Delta_{\bm k}({\bm q})$ for the Cooper pairs with non-zero momentum ${\bm q}$,
\begin{equation}\label{deltas}
\Delta_{\bm k}({\bm q})=-\frac{i}{N}\sum_{{\bm a}, {\bm R}}\Delta_{\bm a}({\bm R})\,e^{-i{\bm q}{\bm R}}\sin{\bm k}{\bm a}=-i\sum_{{\bm a}}\Delta_{\bm a}({\bm q})\,\sin{\bm k}{\bm a}=-2i\left[\Delta_x({\bm q})\sin k_x +\Delta_y({\bm q})\sin k_y\right]\,.
\end{equation}
is a linear combination of $\sin k_x$ and $\sin k_y$ basic functions (we measure distances in the lattice spacing, i.e. $|{\bm a}|=1$). The both wavevectors ${\bm k}$ and ${\bm q}$ run over Brillouin zone, Figure {\ref{fig:bz}}. The spectrum of non-interacting electrons is given by $\xi_{\bm k}=-\cos k_x-\cos k_y -\mu$ with $\mu$ being the chemical potential. All energies are dimensionless, they are measured in terms of the half bandwidth. 

\subsection*{Zero momentum Cooper pairs}
Keeping only (imaginary) time independent part of the pairing field for Cooper pairs, $\Delta_{\bm k}({\bm q}=0)\equiv\Delta_{\bm k}$, one derives the action,
\begin{align}%\nonumber
	S=\frac{1}{2}\sum_{{\bm a}}\frac{|\Delta_{\bm a}({\bm R})|^2}{J_{\bm a}}
	+\frac{1}{2}\sum_{{\bf k}}{\bar\Psi}_{\bm k}\left[\partial_{\tau}+H_{\bm k}\right]\Psi_{\bm k}\,,\quad
	\Psi^{\mathrm T}_{\bm k}=\left(\psi_{\bm k}, {\bar\psi}_{-\bm k}\right)\,,
\end{align}
that contains Bogolyubov-deGennes (BdG) Hamiltonian:
\begin{equation}
H_{\bm k}=
\left(
\begin{array}{cc}
\xi_{\bm k} & \Delta_{\bm k} \\
{\bar\Delta}_{\bm k} & -\xi_{\bm k}  \\
\end{array}
\right)={\bm N}_{\bm k}\,{\bm \sigma}\,,\quad
{\bm N}_{\bm k}=\left(Re\Delta_{\bm k},-Im\Delta_{\bm k}, \xi_{\bm k}\right)\,.
\end{equation}
One can integrate out fermions to obtain the following free energy (per unit cell) for the superconducting field,
\begin{equation}%\label{F0}
F_0=\frac{1}{J}\,\sum_{\mu}\,|\Delta_{\mu}|^2-\frac{T}{N}\sum_{{\bm k}}\ln\cosh\beta E_{\bm k}/2\,,\quad E_{\bf k}=\sqrt{\xi_{\bf k}^2+|\Delta_{\bf k}|^2}\,, \quad
\Delta_{\bm k}=-2i\sum_{\mu}\,\Delta_{\mu}\sin k_{\mu}\,,
\end{equation} 
where $\beta=T^{-1}$ is the inverse temperature, $E_{\bm k}$ is the energy of quasiparticle excitations, and notations $\Delta_{\mu}\equiv\Delta_{{\bm a}_{\mu}}$ with $\mu=x,y$ used. The minimum of the free energy Eq. (\ref{fe}) is achieved \cite{K} at $\Delta_x\pm i\Delta_y=0$. The BdG Hamiltonian is then parametrised with the unit Bloch vector ${\bm n}_{\bm k}={\bm N}_{\bm k}/E_{\bm k}=
{\bm N}_{\bm k}=\left(|\Delta|\sin k_y,\pm|\Delta|\sin k_x, \xi_{\bm k}\right)$.
We have fixed the gauge by requiring ${\mathrm arg}\,{\Delta}_x=0$. This Bloch vector parametrise $p_x\pm i p_y$ superconductor with broken time-reversal symmetry. These are topologically nontrivial phases with the Chern number
\begin{equation}
C_{\pm}=  \int _ { \mathrm { BZ } }\frac { d k _ { x } d k _ { y}  } { 4 \pi  } \,\hat { \mathbf { n } } \cdot \partial _ { k_x} \hat { \mathbf { n } } \times \partial _ { k_y } \hat { \mathbf { n } }
=\pm\,{\mathrm {sign}}\,\mu\,.
\end{equation}
The nonzero net momentum coupling field cannot be treated exactly as it was done above for zero-momentum Cooper pairs and we will now perform analysis of the vicinity of the transition where quadratic approximation in the pairing field is sufficient to identify most dangerous instability.

\subsection*{Cooper pairs with arbitrary net momentum}
Keeping the pairing field $\Delta_{\bm k}({\bm q})$ that describes possible instability of the Cooper pairs with nonzero net momentum ${\bm q}$, one derives the free energy per unit cell near $T_c$, Eq. (\ref{fe}). The anomalous polarisation bubble convoluted with vertex functions,
	\begin{eqnarray}
	\Pi_{\mu\nu}({\bm q})=\frac{1}{N}\sum_{\bm k}\,\sin k_{\mu}\sin k_{\nu}\,\frac{\tanh\beta\xi_{{\bm k}+{\bm q}/2}/2+\tanh\beta\xi_{{\bm k}-{\bm q}/2}/2}{\xi_{{\bm k}+{\bm q}/2}+\xi_{{\bm k}-{\bm q}/2}}\,,
	\end{eqnarray}
defines fluctuations of two fields $\Delta_{x}(\bm q)$ and $\Delta_{y}(\bm q)$, Eq. (\ref{deltas}), with eigenvalues $\Lambda_{\pm}$:
	\begin{equation}\label{Epm}
	\Lambda_{\pm}=J^{-1}+E_{\pm}\,,\quad E_{\pm}=-\Pi_+\,\pm \sqrt{\Pi_-^2+\Pi^2_{\perp}}\,,\qquad
	\Pi_{\pm}=\frac{1}{2}\,\left[\Pi_{11}\pm\Pi_{22}\right]\,,\quad
	\Pi_{\perp}=\Pi_{12}=\Pi_{21}\,.
	\end{equation}
Superconducting instability emerges when the lowest of two branches, $\Lambda_-({\bm q})$, touches zero at some ${\bm q}_c$ as one lowers the temperature:
\begin{equation}
\Lambda_-({\bm q}_0, T_c)=0\,,\qquad \Lambda_{-}=J^{-1}+E_-({\bm q})\,,\quad E_-({\bm q})=-\Pi_+-\sqrt{\Pi_-^2+\Pi^2_{\perp}}\,.
\end{equation}
Our numerical analyses, Figure (\ref{fig:mu01}), shows that in a generic case $\mu\neq 0$ the minimum of $E_-({\bm q})$ is always (for any temperature and chemical potential) located at ${\bm q}=0$. The essential feature of the electron spectrum in the normal phase is that it has mirror symmetries: $\xi_{m{\bm k}}=\xi_{\bm k}$  where $m$ stands for a mirror reflection of the wavevector ${\bm k}$ across vertical and horizontal or diagonal lines. The former imposes $\Pi_-({\bm q}=0)=$ while the latter requires $\Pi_{\perp}({\bm q}=0)$. Due to these facts, both $\pm$ modes are degenerate at ${\bm q}=0$ and superconducting order parameter, which corresponds to zero net momentum of Cooper pairs, is an arbitrary linear combination $\Delta_{\bm k}({\bm q}=0)=-2i(\Delta_1\sin k_x+\Delta_2\sin k_y)$ with no relation between coefficients $\Delta_x$ and $\Delta_y$. This relation will emerge if further expansion (quartic terms) in pairing fields is performed. This will produce restriction \cite{K} on the mean values $\Delta_x\pm i\,\Delta_y=0$.

\subsection*{Half filling}
At the half filling $\mu=0$, there are two lines $q_x=\pm q_y$, see Figure (\ref{fig:mu}), where the minimum of $E_-({\bm q})$ is achieved. The fact that value of $E_-({\bm q})$ is constant on diagonals ${\bm q}={\bm q}_d\equiv (1,\pm 1)\,q_x$ of the Cooper pairs Brillouin zone can be shown analytically. This fact means binding of electrons with momentum ${\bm k}$ and ${\bm q}_d-{\bm k}$ and simultaneous instability of a continuum of pairs with net momentum ${\bm q}$.
To describe transition between superconducting states with Chern numbers $C=1$ and $C=-1$, we may introduce parameter that 
\begin{equation}
\delta E (\mu)=E_-({\bm q}={\bm\pi}; \mu) - E_-({\bm q}=0; \mu)\,,
\end{equation}	
where ${\bm\pi}=(\pi,\pi)$ is the momentum of the corner of the Cooper pairs Brillouin zone. This parameter vanishes at the transition $\mu=0$ between two topological superconducting phases with opposite Chern numbers, see Figure (\ref{fig:mu}).	
	
\bibliography{tsc}

	\section*{Acknowledgements}
	
This work was supported by the Leverhulme Trust Grants RPG-2016-044 (IVY).

\end{document}